
\input harvmac
\input epsf
\def\tr{{\rm Tr}\,}
\def\Jb{\overline J}
\def\Jt{\widetilde J}
\def\zetab{\bar\zeta}

\def\const{\rm const.}
\def\frac#1#2{{#1 \over #2}}
\def\putfig#1#2{\midinsert\bigskip\bigskip
 \centerline{ \epsfbox{#1} }\vskip 0.25truein
 \centerline{\vtop{\hsize=4.5truein\tenpoint
 \baselineskip=14pt plus 2pt minus 1pt
 \noindent #2 }}\smallskip\endinsert}
\def\head#1{\bigbreak\bigskip\noindent{\bf #1}\par\nobreak\medskip\nobreak}
\def\subhead#1{\bigbreak\noindent{\it #1}\par\nobreak\smallskip\nobreak}
\lref\NIEN{B.~Nienhuis, Phys. Rev. Lett. 49, 1062, 1982;
 and `Coulomb Gas Methods in Two-Dimensional Phase Transitions',
 in {\sl Phase Transitions and Critical Phenomena}, C.~Domb and
 J.~L.~Lebowitz, eds. (Academic, New York, 1986), Vol.~11.}
\lref\JCLH{J.~L.~Cardy, `Conformal Invariance and Statistical
 Mechanics', in {\sl Fields, Strings and Critical Behavior},
 proceedings of Les Houches Summer School in Theoretical
 Physics, 1988 (North-Holland, 1990).}
\lref\DUP{B.~Duplantier, `Conformal Invariance and Self-Avoiding Walks',
 in {\sl Fields, Strings and Critical Behavior},
 proceedings of Les Houches Summer School in Theoretical
 Physics, 1988 (North-Holland, 1990).}
\lref\CG{J.~L.~Cardy and A.~J.~Guttmann, `Universal Amplitude
 Combinations for Self-Avoiding Walks,
 Polygons and Trails', J. Phys. A26, 2485, 1993.}
\lref\JCPRL{J.~L.~Cardy, `Mean Area of Self-Avoiding Loops',
 Phys. Rev. Lett. 72, 1580, 1994.}

\Title{}{Geometrical Properties of Loops and Cluster Boundaries}

\centerline{\authorfont John Cardy\footnote{$^\dagger$}{cardy@thphys.ox.ac.uk}}
\bigskip
\centerline{\it All Souls College}
\centerline{and}\centerline{\it Department of Physics}
\centerline{\it Theoretical Physics}
\centerline{\it 1 Keble Road}
\centerline{\it Oxford OX1 3NP, UK}

\vskip .3in
We discuss how the statistical properties of the area and radius of
gyration of single self-avoiding loops, and of Ising and percolation
cluster boundaries, may be calculated using ideas of two-dimensional
field theory. For cluster boundaries, we show that almost all loops
have area $C\ln L+O(1)$, where $L$ is the size of the system, and $C$
is a calculable constant. We also compute the universal ratios
$\langle A\rangle_\ell/\langle R^2\rangle_\ell$ of the area to the
squared radius of gyration of loops of a given large perimeter $\ell$.

\bigskip
\noindent{Presented at Les Houches Summer School on {\it Fluctuating
Geometries in Statistical Mechanics and Field Theory\/}, 2 Aug -- 9 Sep, 1994.}

\Date{9/94}

\head{Introduction.}

If we look at a typical configuration in a Monte Carlo simulation of the
2D Ising model at its critical point, we see a scale-invariant distribution of
structures. If we focus on the boundaries of the clusters, there are some
small ones, but others seem to stretch across the system. What is the
statistics of the size and shape of these objects? This is a geometric
question, not simply related to the conventional correlation functions of
the Ising model.
In fact, if we study typical configurations of the same model at temperatures
$T>T_c$, although it exhibits a finite correlation length in the
conventional sense, the geometrical properties of the cluster boundaries
appear to be very little different. As we shall show in detail, this is
indeed true all the way up to infinite temperature.

We shall discuss two useful measures of the size and shape of these objects,
which, it turns out, may be calculated exactly for large clusters. The first
is the area contained within a cluster boundary, which in general forms a
closed loop. The second is its radius of gyration, defined by
$R^2=(1/2\ell^2)\sum_{r_1,r_2}(r_1-r_2)^2$, where the sum is over pairs of
points on the perimeter, and $\ell$ is the number of such points. As well as
the Ising model for $T=T_c$ and $T>T_c$, we shall also discuss the boundaries
of percolation clusters at and near the percolation threshold $p=p_c$, and
the problem of single self-avoiding loops in the plane. These all turn out
to be special cases of an $O(n)$ model, defined as follows. Consider a
honeycomb lattice (this is for convenience in making a clean definition of
cluster boundaries, but similar results may be obtained for the square
lattice.)
At each site $r$ place an $O(n)$ spin with components $s_a(r)$, where
$a=(1,2,\ldots,n)$, normalised so that $\tr s_a(r)s_b(r')=\delta_{ab}
\delta_{rr'}$. The partition function is
$$
Z_{O(n)}=\tr\prod_{\rm links}\left(1+x\sum_as_a(r)s_a(r')\right)
$$
For small $x$ we may expand this as a sum over configurations of bonds
on the lattice, with a weight $x$ for each bond which is present. After taking
the trace, the only configurations which survive (see Fig.~1)
consist of self-avoiding
closed loops, with a factor $n$ that arises on taking the trace over the
last spin in each loop. Thus
\eqn\ON{
Z_{O(n)}=\sum_{\rm loop\ configurations}
x^{\rm total\ length}n^{\rm number\ of\ loops}
}
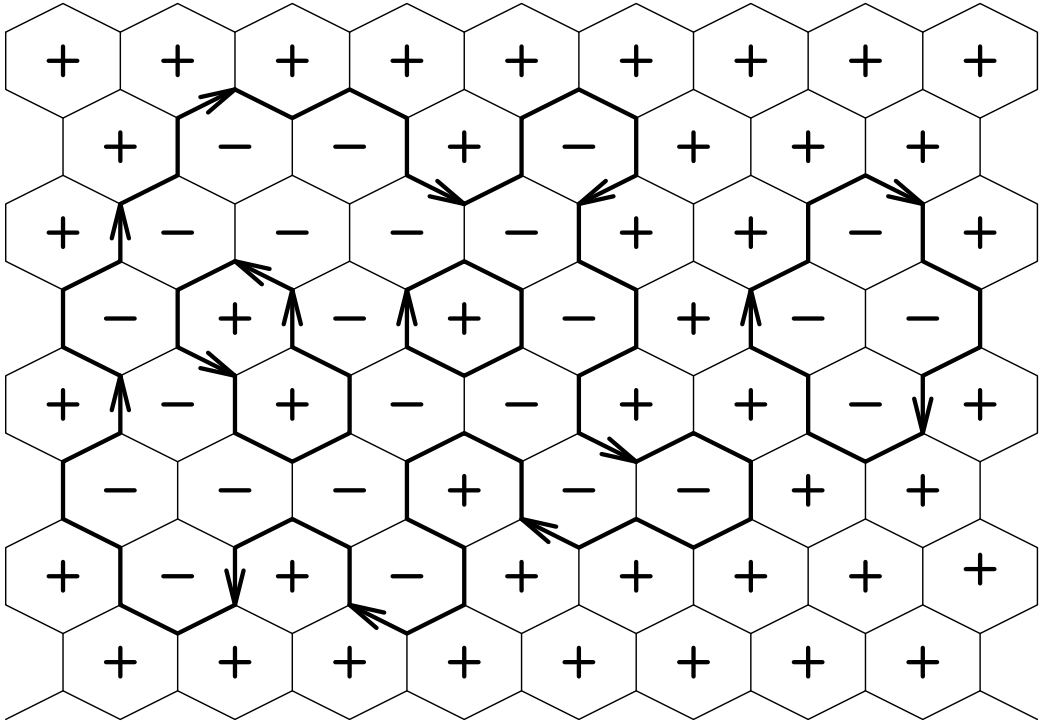
\putfig{dw.eps}{Figure 1. Loop gas on the honeycomb lattice, and
Ising clusters on the dual triangular lattice.}
The advantage of the honeycomb lattice is that it has coordination number
three, so that loops never touch at the vertices.
This is not of course the usual partition function for the lattice
$O(n)$ nonlinear sigma model, which is $Z=\tr\exp(\beta\sum s(r)\cdot
s(r'))$. Nevertheless, they agree at high temperatures, when $x\sim\beta$,
and, for the Ising case $n=1$, they agree exactly with $x=\tanh\beta$.
Unlike the non-linear model, \ON\ makes sense for all $n$, and it has
a continuous transition in two dimensions for $n\leq2$.
The phase diagram is best illustrated by the schematic renormalisation group
flows in Fig.~2. There is a conventional unstable fixed point at
$x=x_{c1}(n)$, but the whole low-temperature phase is also critical in
general, being attracted into a stable non-trivial fixed point at
$x=x_{c2}(n)$. For the case $n=1$, $x_{c2}=1$, corresponding to a
conventional $T=0$ fixed point.
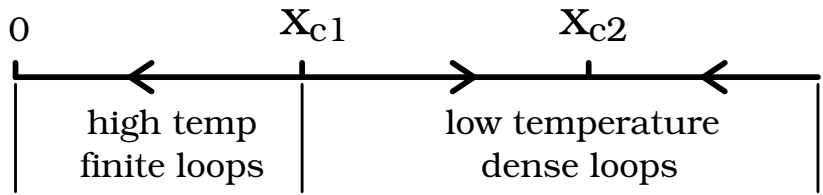
\putfig{xrg.eps}{Figure 2. Schematic RG flows and phase diagram of
the $O(n)$ model as a function of $x$.}
The relation of this model for $n=1$ to cluster boundaries is through
duality. The loops in \ON\ may be thought of as the domain walls of an
Ising model on the dual, triangular lattice. When the original model is
at $x=x_{c1}$, the dual spins are also critical, while $x=x_{c2}=1$
corresponds to the dual spins being at infinite temperature. Since the
whole phase $x>x_{c1}$ is attracted into this fixed point, so is the
whole high-temperature phase of the dual model. Thus the universal large
distance properties of its cluster boundaries should be the same as those
at infinite $T$.

This infinite temperature Ising model on the triangular lattice may also
be interpreted as a site percolation problem, where sites are occupied
($\equiv$ spin up) with probability $\frac12$. This is precisely the
percolation threshold for the site percolation problem on this lattice,
so these cluster boundaries may also be thought of as those of critical
percolation clusters, and their universal properties should apply to
critical site and bond percolation clusters on any lattice.

\head{Area of loops.}

In two dimensions, any gauge field is confining, and as a result, Wilson
loops obey an exact area law. One way, therefore, to express the area
inside a simple closed curve is to regard it as a Wilson loop. To be
explicit, consider a unit current $J_\mu$ flowing around the loop in its
direction of orientation. Then the area is
\eqn\A{
A_{\rm loop}=\int\int G_{\mu\nu}(r-r')J_\mu(r)J_\nu(r')d^2rd^2r'
}
where $G_{\mu\nu}=\langle A_\mu(r)A_\nu(r')\rangle$
is the gauge field propagator. It is simplest to take
the gauge group to be $U(1)$ , so that, in the gauge $A_x=0$, the area
is
$$
A_{\rm loop}=-\frac12\int\int|x-x'|\delta(y-y')J_y(x,y)J_y(x',y')d^2rd^2r'
$$
which simply corresponds to a decomposition of the interior of the loop
into strips of length $|x-x'|$ and width $\delta y=\delta y'$.
Although this is written in a continuum
notation, in fact this is valid even on the lattice if we imagine each
link slightly broadened so as to form a narrow strip, with a current
density $J_\mu$ such that there is unit flux along each link. This
makes the transition to the continuum limit particularly clean and
unambiguous.

What is this current $J_\mu$? This is most easily understood by thinking
of the $O(n)$ model as a {\it complex} $O(n/2)$ model, with complex
spins $S_a=s_a+is_{a+1}$, and partition
function $Z=\tr\prod(1+xS^*\cdot S+{\rm c.c.})$. $J_\mu$ is then just
the $U(1)$ current, proportional to $(1/2i)(S^*\cdot\partial S-
{\rm c.c.})$, whose integral generates the symmetry
$S_a(r)\to e^{i\theta}S_a(r)$. If we now sum over all loops and take the
expectation value, we get
$$
n\langle A\rangle=-\frac12\int\int|x-x'|\langle J_y(x,y)J_y(x',y')
\rangle d^2rd^2r'
$$
where the $\langle JJ\rangle$ correlation function is evaluated in the
$O(n)$ model, and $A$ is the total area of {\it all} loops. Note that
contributions to $\langle J_\mu(r)J_\nu(r')\rangle$ cancel in the sum
over orientations when $r$ and $r'$ are on different loops. If we use
translational invariance, we finally arrive at
\eqn\AF{
\langle A\rangle=-\frac1{2n}{\cal A}\int_{-\infty}^\infty
|x|\langle J_y(x,0)J_y(0,0)\rangle dx
}
where $\cal A$ is the total area of the lattice.

At its critical points, the $O(n)$ model, on distance scales much larger
than the lattice spacing $a$ is supposed to be described by a conformal
field theory. $J_\mu$ is conserved current in this theory with scaling
dimension 1, and, by
scale invariance, rotational invariance, and conservation, its
two-point function must have the form
\eqn\JJ{
\langle J_\mu(r)J_\nu(0)\rangle=k(n)\left({r_\mu r_\nu-\frac12r^2g_{
\mu\nu}\over r^4}\right)
}
which is unique apart from the universal number $k$ which characterises
the theory (and the choice of $U(1)$ current.) [Note that $k$ is
sometimes called the chiral anomaly, because if we define the left-
and right-moving components of the current to be $J_{L,R}=
J_1\pm iJ_2$, strict current conservation requires the existence of
a contact term $J_L(r)J_R(r')\propto k\delta^{(2)}(r-r')$.] In our case
we see that the correlation function in \AF\ behaves like $1/x^2$ so
the integral is logarithmically divergent at both short and large
distances. Short distances are presumably cut off at the order of
the lattice spacing, since the continuum description breaks down there,
but the infrared divergence is real, and indicates that the average area
has no thermodynamic limit. Instead, if we evaluate this in a finite
system of linear size $\sim L$, we should find
$$
\langle A\rangle/{\cal A}\sim (k(n)/2n)\ln(L/a)
$$
The right hand side is, of course, much greater than unity for large
$L$. This can happen because there are loops with loops, and thus some
portions of the total area are counted many times in $\langle A\rangle$.

In the same way, higher moments of $A$ are given by integrals over
higher order correlations of the current $J$. Fortunately, these are
simple to write down, since it is an elementary fact of conformal field
theory that the connected pieces of such correlation functions all
vanish. The way to prove this is to consider the holomorphic and
antiholomorphic components $J=J_1-iJ_2$ and $\Jb=J_1+iJ_2$. Current
conservation implies that a
correlation function
$$
\langle J(z_1)J(z_2)\ldots\Jb(\zetab_1)\Jb(\zetab_2)\ldots\rangle
$$
depends only holomorphically on the $\{z_j\}$ and antiholomorphically
on the $\{\zetab_j\})$, so that it is entirely determined by its
singularities as these points approach each other.
These are given by the operator product expansion
$$
J(z_1)J(z_2)\sim{k/4\over(z_1-z_2)^2}+{\rm regular\ terms}
$$
with similar relations for $\Jb$. The important point is that, for a
$U(1)$ current, there is no term $O((z_1-z_2)^{-1})$. The consequence is that,
if we subtract off the disconnected part of the correlation function,
the remainder has no singularities, and falls off sufficiently fast at
infinity that it must vanish identically.

Thus, apart from the value of $k$, the correlations of a $U(1)$ current
are the same in any conformal field theory. We are therefore free to use
a free field representation, which, for convenience, we take to be
$J_\mu=\sqrt k\epsilon_{\mu\nu}\partial_\nu\phi$, where
$\phi(r)$ is a free scalar field with action $S_G=(1/2\pi)\int
(\partial\phi)^2d^2r$. Since the loops do not cross the boundary, the
normal component of $J$ should vanish there, corresponding to choosing
Dirichlet conditions $\phi=0$ on the boundary. It is also more
convenient to write \A\ in a covariant gauge, for which
$\partial_\mu G_{\mu\nu}=0$ and $\partial^2G_{\mu\nu}=-\delta^{(2)}(r-r')$.
It is then readily checked, using Stokes' theorem, that \A\ correctly
gives the area.

To generate the higher cumulants of $A$ it is convenient to consider its
generating function, given in the free field representation by
$$
\langle e^{-uA}\rangle=
\langle e^{-u(k/n)\int\int G_{\mu\nu}(r-r')\epsilon_{\mu\lambda}
\epsilon_{\nu\sigma}\partial_\lambda\partial'_\sigma\phi(r)\phi(r')
d^2rd^2r'}\rangle
$$
Integrating by parts and using the above properties of $G_{\mu\nu}$ then
gives a mass term for the $\phi$ field, proportional to $(k/n)u$, so
that the result is simply a ratio of two Gaussian integrals
\eqn\GF{
\langle e^{-uA}\rangle=
{{\rm det}(-\partial^2+(2\pi ku/n))^{-1/2}\over
{\rm det}(-\partial^2)^{-1/2}}
=e^{-\frac12\sum_m\ln\left(1+{(2\pi ku/n)L^2\over\lambda_m}\right)}
}
where the $\lambda_m/L^2$ are the eigenvalues of $-({\rm laplacian})$,
with Dirichlet boundary conditions.

If we expand \GF\ to first order in $u$, we recover our previous
result
$$
\langle A\rangle=(\pi k/n)L^2\sum_m{1\over\lambda_m}
$$
since the sum has a divergence like $(1/2\pi)^2\int d^2p/p^2\sim
(1/2\pi)\ln(L/a)$. However, the higher cumulants behave as
$$
\langle A^p\rangle_c\propto L^{2p}\sum_m{1\over\lambda_m^p}
$$
and this sum is convergent (but dependent on the shape of the region)
for $p>1$.

This result implies that $A/{\cal A}$ has mean value $O(\ln L)$, but
with fluctuations $O(1)$. However, we are more interested in the area
per loop, given by $A/N_l$, where $N_l$ is the number of loops. Since
this is conjugate to the fugacity $n$, its cumulants are given by
$$
\eqalign{
\langle N_l\rangle&=(n\partial/\partial n)\ln Z_{O(n)}\cr
\langle N_l^2\rangle-\langle N_l\rangle^2
&=(n\partial/\partial n)^2\ln Z_{O(n)}\cr}
$$
and so on. Each of these is an extensive quantity, proportional to
$\cal A$, even at the critical point. (In fact, they are calculable
exactly thanks to a series of mappings to the 6-vertex model.)
Hence we see that $N_l$ is of order $L^2\sim{\cal A}$, with fluctuations
of order $L$. Therefore in the ratio $A/N_l$ the numerator fluctuates
much more strongly than the denominator, and it is permissible to
replace $\langle A/N_l\rangle$ by $\langle A\rangle/\langle N_l\rangle$.

The final result is that, not only is {\it average} size of a loop
large,
but, with probability one as $L\to\infty$, {\it all} loops have size
$C\ln L+O(1)$. The amplitude $C$ is a ratio of the universal number
$k(n)/2n$ and the non-universal, lattice-dependent, quantity
$\langle N_l\rangle/{\cal A}$. From \GF\ one may also recover the
whole probability distribution $P(A)$ of $A$, in the limit when $a$ and $L$
are large, by inverse Laplace transform. In particular, for
$$A-\langle A\rangle\gg L^2\ ,$$
we find an exponential decay
$$
P(A)\sim e^{-[\lambda_0/(2\pi k/n)](A/L^2)}
$$
where $\lambda_0/L^2$ is the lowest eigenvalue of $-\partial^2$.

The above results for the average area may be generalised away from
the point where the loops are critical. In the percolation problem,
this would correspond to $p\not=p_c$, and, for Ising clusters, to
adding an external magnetic field $H$. In either case the integral in
\AF\ will now behave like $(k/2n)\ln\xi$, where the correlation
length $\xi$ behaves as
$\xi\sim(p_c-p)^{-\frac43}$ for percolation, and for the Ising
model as $H^{-\frac{15}8}$ ($T=T_c$) and $H^{-\frac12}$ ($T>T_c$).

\head{Finite-length loops.}

Although the above calculation told us about the average statistics of
loop areas, it did not tell us about the size or shape of individual
loops. If we pick out one loop and measure its length to be $\ell$, how
does its area and radius of gyration depend on $\ell$?

To study this question we need a generalisation of the $O(n)$ model,
in which the indices $a$ run from 1 to $n+n'$:
$$
Z=\tr\prod_{\rm bonds}\left(1+x\sum_1^ns_a(r)s_a(r')+
x'\sum_{n+1}^{n+n'}s_a(r)s_a(r')\right)
$$
This gives us an expansion in powers of $x$ and $x'$ which is a sum
over configurations with two different kinds of mutually non-overlapping
loops. As $n'\to0$, the term $O(n')$ corresponds to those configurations
with single primed loop:
$$
Z=Z_{O(n)}\left(1+n'{\cal A}\sum_\ell p_\ell {x'}^\ell+O({n'}^2)
\right)
$$
which defines the coefficient $p_\ell$. (As $n\to 0$ this is just the
total number of self-avoiding loops of length $\ell$, weighted equally.)
This model has a critical point at $x=x'=x_c$. Turning on the
perturbation $(x'-x_c)$ breaks the symmetry down to $O(n)\times O(n')$,
and we expect the degrees of freedom with $n<a\leq n'$ to become
massive.

To measure the area of the chosen loop we may use the current
$$J'\sim(1/2i)\sum_{n+1}^{n+n'}(s_a^*\partial s_a-{\rm c.c.})\ ,$$
whose 2-point function is given by $\langle J'J'\rangle=(n'/n)
\langle JJ\rangle$. Then, as before, on taking the expectation value in
this ensemble,
\eqn\AL{
n'\sum_\ell p_\ell\langle A\rangle_l\,{x'}^\ell
=-\frac12\int|x|\langle J'_y(x,0)J'_y(0,0)\rangle dx
}
where $\langle A\rangle_\ell$ is the mean area of loops of length $\ell$
when embedded in the gas of critical loops of the $O(n)$ model. The
integrand once again behaves as $1/x^2$, but it is now cut off at large
distances by the length scale $\xi'$ which is essentially the inverse
mass of the massive modes. This will diverge as $x'\to x_c$ as
$\xi'\sim(x_c-x')^{-\nu'}$, where $\nu'$ is related to the scaling
dimension $X'$ of the perturbation by $\nu'=2/(2-X')$. Thus the right
hand side of \AL\ behaves like $\ln(x_c-x')$ with a calculable
coefficient. From this we discover the large $\ell$ behaviour of the
coefficient of ${x'}^\ell$ on the left hand side:
$$
p_\ell\langle A\rangle_\ell\sim (k(n)/2n)\nu'(n)\ell^{-1}x_c^{-1}
$$
Now $\sum_\ell p_\ell{x'}^\ell$ itself is like a free energy, which,
by hyperscaling, should have a singular part which behaves like
${\xi'}^{-2}\sim (x_c-x')^{2\nu'}$, which gives the large $\ell$
behaviour of $p_\ell$. Putting these results together, we find that
$ p_\ell\sim \ell^{-1-2\nu'}x_c^{-1}$ and
$\langle A\rangle_\ell\sim\ell^{2\nu'}$.

\subhead{Radius of gyration.}

The radius of gyration of a single loop is given by
$R^2=(1/2\ell^2)\sum_{r_1,r_2}(r_1-r_2)^2$.
We need to find a way of counting all loops going through two
chosen bonds at $r_1$ and $r_2$. Let us define the energy density of the
perturbation of the $O(n+n')$ model on the bond at $r$ as
$E'(r)=\sum_{n+1}^{n+n'}s_a\cdot s_a$. Then $E'(r_1)E'(r_2)$
has only contributions from the selected loops. In the limit $n'\to0$,
$r_1$ and $r_2$ must be on the same loop, since there is only one. Hence
$\sum_{r_1,r_2}(r_1-r_2)^2E'(r_1)E'(r_2)$, when inserted into a correlation
function, gives the radius of gyration (squared) of the loop. On
averaging, we thus find that
$$
2n'{\cal A}\sum_\ell p_\ell\ell^2\langle R^2\rangle_\ell
\,{x'}^{\ell-2}=
\sum_{r_1,r_2}\langle (r_1-r_1)^2E'(r_1)E'(r_2)\rangle
\sim {\cal A}\int r^2\langle E'(r)E'(0)\rangle d^2r
$$
where this last correlation function is to be evaluated in the
{\it perturbed} $O(n+n')$ model. This last integral may be evaluated by
a corollary of Zamolodchikov's $c$-theorem.\JCLH

The addition of the term $\sum_r(x'-x_c)E'(r)$ to the action takes the
theory away from the conformal point, and gives rise to a non-zero trace
$\Theta=T^\mu_\mu$ of the stress tensor. In fact, $\Theta=
2\pi{\nu'}^{-1}(x'-x_c)E'(r)$, which may be demonstrated by considering
the response of the theory to a dilatation.\JCLH\ Thus we need
the second moment of 2-point function $\langle\Theta\Theta\rangle$.
A simple way of deriving this is as follows: in two dimensions,
conservation and rotational symmetry force two-point function of the
stress tensor to have the form
$$
\langle T_{\mu\nu}(r)T_{\lambda\sigma}(r')\rangle=
(\partial_\mu\partial_\nu-g_{\mu\nu}\partial^2)
(\partial'_\lambda\partial'_\sigma-g_{\lambda\sigma}{\partial'}^2)
G(r-r')
$$
where $G$ is a scalar function. At short distances, the theory is
equivalent to some conformal field theory with central charge
$c_{UV}$, so that
$$
\langle T_{zz}(z)T_{zz}(z')\rangle\sim{c_{UV}/2\over(z-z')^4}
\sim \partial^2{\partial'}^2G
$$
so $G(r)\sim(c_{UV}/6)\ln r$ as $r\to0$. Similarly, as $r\to\infty$,
$G(r)\sim(c_{IR}/6)\ln r$.
Now $\int r^2\langle\Theta(r)\Theta(0)\rangle d^2r=\int r^2\partial^4
G(r)d^2r$. Integrating by parts twice kills the factor $r^2$, and the
result is the integral of a total derivative, which is not zero because
of the surface terms at $r=0$ and infinity. The final result is that
\eqn\CTH{
c_{UV}-c_{IR}={3\over4\pi}\int r^2\langle\Theta(r)\Theta(0)\rangle d^2r
}

In our case, the UV theory is the $O(n+n')$ model. In the IR, we are
left with the massless modes of the $O(n)$ model. Hence the left hand
side of \CTH\ is $c(n+n')-c(n)\sim n'(dc/dn)$, so that
$$
2\sum_\ell p_\ell\ell^2\langle R^2\rangle_\ell\,{x'}^{\ell-2}
\sim \big(2\pi{\nu'}^{-1}(x'-x_c)\big)^{-2}(4\pi/3)(dc/dn)
$$
This tells us about the large $\ell$ behaviour of $p_\ell\langle
R^2\rangle_\ell$, and, in particular that
$\langle R^2\rangle_\ell\sim \ell^{2\nu'}$.
This implies that a loop of finite (but large) length, immersed in the
gas of other critical loops, has a fractal dimension of ${\nu'}^{-1}$.
The amplitudes in the mean area and $R^2$ are not universal, but
the ratio, in which all dimensionful and cut-off dependent
quantities cancel, is, and we find, eliminating $p_\ell$,
$$
{\langle A\rangle_\ell\over\langle R^2\rangle_\ell}
\sim3\pi{k(n)/n\over(dc/dn)\nu'(n)}
$$

The remarkable thing about this equation is that it relates expectation
values at finite $\ell$, that is, in the non-critical $O(n)\times O(n')$
model, to quantities defined at the conformal points.

\head{Calculation of $k$, $c$ and $\nu'$.}

It turns out that these universal quantities may
be computed using Coulomb gas techniques. These are explained at length
in the review article by Nienhuis\NIEN\ (see also \DUP),
so they will be only summarised here. We first rewrite the partition
function \ON\ of the $O(n)$ model in a form where the Boltzmann weights
are local. This may be done by assigning to each vertex of each oriented
loop a factor $e^{i\chi}$ or $e^{-i\chi}$ according to whether the
oriented walk turns to the left or the right as it passes through the
vertex. In this way, each anticlockwise loop will accumulate a factor
of $e^{6i\chi}$, and each clockwise loop $e^{-6i\chi}$. On summing over
orientations, we may reproduce the factor $n$ for each loop if we choose
$\chi$ such that $n=2\cos6\chi$.

The next step is to write the loop gas as a solid-on-solid (SOS) model.
Define a height variable $\phi(r)$, conventionally normalised so that
$\phi(r)/\pi$ is an integer, at each site of the dual triangular
lattice. There is 1-1 mapping between configurations of oriented loops
and SOS heights as follows: assign $\phi=0$ on the boundary, and
increase (decrease) $\phi$ by $\pi$ each time a loop is crossed which
goes to the left (right). The fact that all loops are closed makes this
a consistent procedure. The resulting Boltzmann weight for each
configuration is now a product of weights for each elementary triangle,
each of which depends only on the {\it differences} of the $\phi$s at
the vertices of the triangle. In the large distance limit, we now assume
that this model renormalises onto one where the discreteness of the
$\phi$s is irrelevant. This may be justified {\it a posteriori.}
Since the continuum limit must be a conformal field theory, depending on
a single scalar field $\phi$, the simplest guess is the Gaussian model
with action $S=(g/4\pi)\int(\partial\phi)^2d^2r$. (Note that this is
different from that introduced to give the free field representation
of $J_\mu$.) The only difficult part of this argument is in assigning
the value of $g$, which determines the various scaling dimensions. The
argument is quite involved\NIEN\ and we quote
only the result $g=1-(6\chi/\pi)$, which corresponds to
$n=-2\cos\pi g$. The correct branches are given by $1\leq g\leq2$,
corresponding to $x=x_{c1}$, and $0\leq g\leq1$, corresponding to
$x=x_{c2}$.

The free field theory is often called a Coulomb gas, because operators
$e^{iq\phi}$ behave like charges of strength $q$:
$\langle e^{iq\phi(r)}e^{-iq\phi(0)}\rangle\sim e^{-(q^2/g)\ln r}$.
However, the SOS model is not completely equivalent to a free field
theory, since if we calculate $\langle e^{-12i\chi\phi/\pi}\rangle$
in the SOS model we find that it is identically equal to 1.
This may be seen to occur to all orders in the expansion in powers of
$x$, the first few terms of which are
$$
\langle e^{-12i\chi\phi/\pi}\rangle=
{1+e^{-12\chi}\cdot e^{6\chi}+e^{12\chi}\cdot e^{-6\chi}+\ldots\over
1+e^{6\chi}+e^{-6\chi}+\ldots}
$$

Since the
only non-zero expectation values in the free field theory are those with
total charge zero (as a consequence of the $U(1)$ symmetry $\phi\to\phi
+\const$), there must be a charge $+12\chi/\pi$ distributed on
the boundary. The only non-zero expectation values are those of products
of operators with total charge $-12\chi/\pi$ in the interior.

How do we identify the current $J_\mu$ of the $O(n)$ model in the
Coulomb gas language? The simplest guess is to take $J_\mu=(1/\pi)
\epsilon_{\mu\nu}\Delta_\nu\phi$, where $\Delta\nu$ is a lattice
derivative, since this will ensure a unit current to the left or right
according to whether $\phi$ steps up or down by $\pi$. However, this
has the wrong charge, and in fact is not correct, as may be seen by
computing its expectation value for a given loop, on summing over
orientations:
$$
\langle J\rangle=1\cdot e^{6i\chi}+(-1)\cdot e^{-6i\chi}\not=0
$$
Instead consider $\Jt_\mu\propto\epsilon_{\mu\nu}\Delta_\nu(
e^{-12i\chi\phi/\pi})$, which does have the correct total charge.
Now
\eqn\VAN{
\langle\Jt\rangle\propto(e^{-12i\chi}-1)e^{6i\chi}
+(e^{12i\chi}-1)e^{-6i\chi}=0
}
as required. However, the 2-point function cannot be
$\langle\Jt\Jt\rangle$, since, once again, this has the wrong total
charge. In fact, what works is $\langle\Jt(r)J(r')\rangle$. When
$r$ and $r'$ are on different loops this vanishes, because we get
the same factor as in \VAN\ on summing over the orientations of the loop
passing through $r$. When they are on the same loop, we get something
proportional to
$$
1\cdot(e^{-12i\chi}-1)\cdot e^{6i\chi}
+(-1)\cdot(e^{12i\chi}-1)\cdot e^{-6i\chi}=-4i\sin6\chi
$$
which may be absorbed into the normalisation of $\Jt$. This is an
example of how the mapping to the Coulomb gas is not at the level of
operators, but of correlation functions. It is now a simple matter
to evaluate $\langle\Jt(r)J(r')\rangle$ in the Gaussian model. It is of
the required form \JJ, with
$$
{k(n)\over n}={1-g\over 2\pi^2g\sin\pi g}
$$

The central charge $c(n)$ of the $O(n)$ model is known by a variety of
methods. A simple way\JCLH\
is to note that it is related to the free energy
of a system when placed in a box of linear size $L$:
$$
F=-\ln Z\sim -(c\chi_E/6)\ln L
$$
where $\chi_E$ is the Euler character. For the free field theory $c=1$,
but the charge on the boundary gives an additional contribution to the
electrostatic energy. Choosing for convenience a disc with $\chi_E=1$,
we then find
$$
{c\over6}={1\over6}-{(6\chi)^2\over g}\qquad{\rm or}\qquad
c=1-{6(g-1)^2\over g}
$$

The final ingredient is $\nu'$, related to the scaling dimension $X'$
of the perturbation $\sum_{n+1}^{n+n'}s_a(r)s_a(r')$.
In the context of
spin systems, this is a symmetry-breaking perturbation inducing a
crossover to lower spin dimensionality, and its analysis is somewhat
subtle. It is first necessary to decompose the perturbation into
irreducible representations of $O(n+n')$, by writing it as
$$
\sum_{n+1}^{n+n'}\left(s_a(r)s_a(r')-{1\over n+n'}\delta_{aa}E(r)
\right)+{n'\over n+n'}E(r)
$$
where $E(r)=\sum_1^{n+n'}s_a(r)s_a(r')$ is the energy density of
the $O(n+n')$ model. The second term contributes merely to a shift in
the critical point $x_c$ when we turn on the perturbation, and is not
important. In the first term it is permissible to set $r'=r$ (since the
terms have the same symmetry) and we notice that each term in the sum
transforms according to the representation of traceless rank 2 tensors.
The scaling dimension is therefore independent of $n'$, and we may use
Nienhuis' result\NIEN\ for the uniaxial case $n'=1$, obtained
by Coulomb gas methods. The result is $\nu'=2g/(1+2g)$. Finally, putting
all the pieces together, we find
$$
{\langle A\rangle_\ell\over\langle R^2\rangle_\ell}\sim
{1+2g\over2(1+g)}\,\pi
$$
This result tells us about the compactness of the loops. If all loops
were circular, the ratio would be $\pi$. Compact loops would have a
ratio larger than this. In fact, we see for all $g$ that this ratio
lies between $\frac12\pi$ and $\pi$. For example, for Ising clusters
avove $T_c$ ($g=\frac23$) we find $\frac7{10}\pi$,
for Ising clusters at $T_c$ ($g=\frac43$) the
slightly greater ratio $\frac{11}{14}\pi$, and for single self-avoiding
loops\JCPRL\ ($g=\frac32$), the value $\frac45\pi\approx2.513$.
This latter value has been checked numerically by extrapolating the
results of exact enumerations for finite $\ell$. The result is
$2.515\ldots$ \CG, which agrees well with the analytic result.

\listrefs\bye